\documentclass[11pt,a4paper,oneside]{article}
\usepackage{graphicx,color}
\usepackage[margin=2cm]{geometry}
\usepackage[T1]{fontenc}
\usepackage{ae,aecompl}
\usepackage{graphicx}	
\usepackage{amsmath}	
\usepackage{amssymb}
\usepackage{lipsum}
\usepackage{authblk}
	
\begin{document}
\title{Multiwavelength studies of X-ray selected extragalactic sample}
\author{A. M. Mickaelian$^{1}$, G. M. Paronyan$^{1}$, G. S. Harutyunyan$^{2}$, H. V. Abrahamyan$^{1}$,\\ M. V. Gyulzadyan$^{1}$}
\affil{
\scriptsize{1.NAS RA V. Ambartsumian Byurakan Astrophysical Observatory (BAO),\\
Byurakan 0213, Aragatzotn Province, Armenia\\
2.Leibnitz Institute for Astrophysics, An der Sternwarte 16, 14482 Potsdam, Germany\\
E-mail: aregmick@yahoo.com}}
\maketitle

\begin{abstract}
The joint catalogue of Active Galactic Nuclei selected from optical identifications of X-ray sources was created as a combination of two samples: Hamburg-ROSAT Catalogue (HRC) and Byurakan-Hamburg-ROSAT Catalogue (BHRC). Both are based on optical identifications of X-ray sources from ROSAT catalogues using low-dispersion spectra of Hamburg Quasar Survey (HQS). However, HRC and BHRC contain a number of misidentifications and using the recent optical and multiwavelength (MW) catalogues we have revised both samples excluding false AGN and adding new genuine ones. Thus a new large homogeneous complete sample of 4253 X-ray selected AGN was created. 3352 of them are listed in the Catalogue of QSOs and Active Galaxies and 387 also are in Roma Multifrequency Catalogue of Blazars. 901 candidate AGN are subject for further study. We classified 173 of these objects using their SDSS DR12 spectra. Following activity types were revealed: 61 AGN, 21 HII galaxies, 12 emission-line galaxies without definite type, 71 absorption-line galaxies, 2 stars, and 6 were classified as $"$Unknown$"$.A special emphasis is made on narrow-line Sy1.0-Sy1.5 galaxies and QSOs, as many of them have soft X-ray, strong FeII lines, and relatively narrow lines coming from BLR ($"$narrow broad lines$"$). As a result, the sample of genuine AGN was enlarged to 3413 objects. We have retrieved MW data from recent catalogues and carried out statistical investigations for the whole AGN sample. An attempt to find connections between fluxes in different bands for different types of sources, and identify their characteristics thus confirming candidate AGNs have been carried out. We have analyzed X-ray properties of these sources to find a limit between normal galaxies and X-ray AGN.\\

\end{abstract}

\section{Introduction}
During the recent decades, several X-ray surveys were carried out and several catalogues of X-ray sources were produced, including most recent XMM-Newton and Chandra sources. However, until now ROSAT remains the only enough deep all-sky survey. ROSAT catalogues contain thousands of interesting objects, including AGN (especially many blazars), hot subdwarfs (sb), white dwarfs (WD), cataclysmic variables (CV), carbon (C) stars, etc. ROSAT sources are listed in two main catalogues: \textbf{Bright Source Catalogue (BSC}, Voges et al. 1999) and \textbf{Faint Source Catalogue (FSC}, Voges et al. 2000). ROSAT BSC contains 18,806 sources with CR $\geq$0.05 cts$\cdot$s$^{-1}$ in the 0.1-2.4 keV energy range, while FSC contains 105,924 fainter sources (CR $\leq$0.05 cts$\cdot$s$^{-1}$). Thus, in total 124,730 sources are available.
Among the identification works, the ROSAT Bright Sources (RBS, Schwope et al. 2000) is well-known; 2012 BSC sources with CR $\geq$ 0.20 and $|$b$|$>30$^{0}$ have been optically identified. However, most of the identified sources come from the Hamburg Quasar Survey (HQS, Hagen et al. 1995). Its low-dispersion spectra allow a preliminary classification of objects into a number of types, giving possibility to make up subsamples of objects for further studies. Two main projects have been carried out: \textbf{Hamburg ROSAT Catalogue (HRC}, Zickgraf et al. 2003) and \textbf{Byurakan-Hamburg ROSAT Catalogue (BHRC}, Mickaelian et al. 2006). Both have been carried out by means of Hamburg Quasar Survey (HQS; Hagen et al. 1995) low-dispersion spectra. HRC is based on ROSAT BSC and contains 5341 sources at $|$b$|$>20$^{0}$  and $\delta$>0$^{0}$.\\
\textbf{BHRC} used brighter sources of ROSAT Faint Source Catalogue (FSC), thus extending the sample to count rates (CR) of photons $\geq$ 0.04 ct/s in the area with galactic latitudes $|$b$|$$\geq$20$^{0}$ and declinations $\delta$$\geq$0$^{0}$ (the area of HQS).It contains 2791 sources. Main advantages of BHRC compared to HRC were the use of DSS1 $\&$ DSS2 red/blue/IR images (which allowed revealing faint objects, more details, extension, proper motions, variability, and accurate positions), the use of 2D HQS images (for quicker selection and better classification), the refinement of classification (identifying QSOs/AGN from SED), introducing new class of interacting galaxies (pairs and groups), search in real 3$\sigma$ positional errors for each source compared to standard search radius in HRC, cross-correlation with MAPS and USNO-B, associations in NED, SIMBAD, catalogues of AGN, WDs, and CVs. Out of 2791 sources, only 95 have no identification; 2696 are identified with 3187 objects (some X-ray sources are identified with double or multiple object). \\
Thus, out of 124,730 ROSAT sources, only $\thicksim$10,000 have optically identifications and ROSAT catalogues still remain as a rich source for new interesting objects.

\section{Merged HRC/BHRC and its extragalactic sample}
To improve the accuracy of measurements and collect more data for HRC sources, we have carried out similar to BHRC studies for HRC, thus establishing homogeneous joint sample of ROSAT identifications. A \textbf{Joint HRC/BHRC Catalogue} has been created by merging HRC and BHRC, altogether 8132 sources (Paronyan $\&$ Mickaelian 2015), the largest sample of identified ROSAT sources. Together with combining data from the two catalogues, we have checked and added many new ones, as well as corrected many errors and misidentifications.\\
Based on HRC/BHRC sample, we have established the types of objects that appear to be X-ray sources. Main types of optical counterparts are: bright (typically spiral) galaxies, blue galaxies (AGN candidates), interacting galaxies (pairs $\&$ groups, those with red colours?), blue compact galaxies (BCG) that appear to be QSOs and Seyferts, blue stellar objects (BSOs) that appear to be QSOs or white dwarfs (WD), bright stars (mostly FG type, including pairs and triples), K-M stars (including binary K-stars), WDs and subdwarfs, cataclysmic variables (CVs). \textbf{Table 1} gives the distribution of the identified objects by types in HRC, BHRC and Joint HRC/BHRC catalogues.\\
\begin{table}[h]
\begin{center}
 \caption{Distribution of objects by types in HRC, BHRC and Joint HRC/BHRC.}
  \begin{tabular}{| l | c | c | c | c | c | c |}
    \hline
  \textbf{Types of objects} & \textbf{HRC} & \textbf{$\%$} & \textbf{BHRC} & \textbf{$\%$} & \textbf{Joint HRC/BHRC}   & \textbf{$\%$} \\ \hline
AGN	& 2215 & 41.5 & 1521 & 54.5 & 4253 & 52.3  \\ \hline
Galaxy & 238 & 4.5 & 114 & 4.1 & 492 & 6.1  \\ \hline
galaxy cluster & 262 & 4.9 & 124 & 4.4 & - & -   \\ \hline
white dwarf & 45 & 0.8 & 28 & 1.0 & 62 & 0.8  \\ \hline
cataclysmic variable & 26 &	0.5 & 11 & 0.4 & 35 & 0.4  \\ \hline
AFG star & 45 & 0.8 & 476 & 17.1 & 327 & 4.0  \\ \hline
K star & 141 & 2.6 & 262 & 9.4 & 310 & 3.8  \\ \hline
M / C star & 197 & 3.7 & 61 & 2.2 & 147 & 1.8  \\ \hline
bright stars & 1219 & 22.8 & 93 & 3.3 & 919 & 11.3  \\ \hline
unidentified & 604 & 11.3 & 6 & 0.2 & 1143 & 14.1  \\ \hline
empty field & 155 & 2.9 & 95 & 3.4 & 250 & 3.1  \\ \hline
no spectrum & 194 & 3.6 & - & - & 194 & 2.4 \\ \hline
\textbf{All} & \textbf{5341}  & \textbf{100.0} & \textbf{2791} & \textbf{100.0} & \textbf{8132} & \textbf{100.0} \\ \hline
  \end{tabular}
\end{center}
\end{table}
In BHRC, the number of objects in most of types are similar to those in HRC, however, we have tried to classify most of the bright stars as well, and as a result, we have very small number of bright stars left (3.3$\%$ compared to 22.8$\%$ in HRC) and a lot of classified stars (30.1$\%$ compared to 8.5$\%$ in HRC). In addition, a larger number of associations has been classified, and most of the analogues of unidentified objects in HRC turned to be faint AGN in BHRC. Therefore, these numbers (41.5$\%$ in HRC and 54.5$\%$ in BHRC) are quite different. After new verifications and corrections in the Joint HRC/BHRC, we have the final distribution of types in the last two columns. Thus, we have 4253 AGN and their candidates, 492 galaxies, 1800 stars and 1587 unidentified objects in the whole sample. Stars are subject for individual studies (Mickaelian et al. 2015). Here we investigate the extragalactic sample.\\

In this paper, we carry out some studies of HRC/BHRC extragalactic sample, including multiwavelength (MW) cross-correlations, statistical investigation, spectroscopic classification based on SDSS DR12 (Alam et al. 2015), etc. Among the 4253 AGN, based on our low-dispersion classification, we have QSOs, blazars, other AGN and blue galaxies, also candidate AGN. All they are strong X-ray emitters, mostly including several definite types of AGN. To understand which types of AGN are expected in our sample, here we characterize most typical extragalactic X-ray sources.\\

\textbf{QSOs} or \textbf{Quasars} have very high luminosities (M$_{abs}$>-23). The optical spectra are similar to those of S1 nuclei, but the narrow lines are generally weaker. There are radio-loud QSOs (quasars or RL QSOs) and radio-quiet QSOs (or RQ QSOs) with a dividing power at P$_{5GHz}$ $\approx$ 10$^{24.7}$ W$\cdot$Hz$^{-1}$. RL QSOs are 5-10$\%$ of the total of QSOs. There is a big gap in radio power between RL and RQ varieties of QSOs. Having a large subsample of X-ray selected QSOs with their MW data, we can study the important question of possible radio and X-ray luminosities correlation. \textbf{OVV} (Optically-Violently Variable) QSOs are similar to BL Lac objects but with normal QSO spectrum. They are radio loud. \textbf{HPQ} or \textbf{HP} (Highly Polarized) Quasars have typical polarization >3$\%$. They are typically combined with OVV quasars as a single class. These objects provide much more fraction of X-ray sources than normal QSOs.\\

\textbf{BL Lac} or \textbf{BLL} (BL Lacertae) type objects. Stellar in appearance with variable, intense and highly polarized continuum. They have strong featureless continuum; no emission or absorption lines deeper than $\sim$2$\%$ are seen in any part of the optical spectrum, or only extremely week absorption and/or emission lines are observed, as a rule at minimum of their very highly variable phase. The weak lines often just appear in the most quiescent stages. So that their redshifts can only be determined from features in spectra of their host galaxies. They show polarization, and are strong radio sources with flat spectrum, as well as strong X-ray sources.\\

\textbf{Blazars} are the combination of two most powerful AGN classes; BLL and OVV/HPQ. These are believed to be objects with a strong relativistically beamed jet in the line of sight. When the angle between the relativistic jet axis and the line of sight is small, the jet is Doppler boosted by a large factor and the whole spectrum (from radio to $\gamma$-ray) is dominated by a compact, highly polarized, highly variable, superluminal, almost featureless continuum, called blazar.\\

\textbf{NLS1, S1n} (Narrow-line Seyfert 1). These were defined by Osterbrock $\&$ Pogge (1985) as soft X-ray sources, having narrow permitted lines only slightly broader than the forbidden ones. Many FeI, FeII, FeIII, and often strong [FeVII] and [FeX] emission lines are present, unlike what is seen in Seyfert 2s. The ratio [OIII]5007/H$_{\beta}$ < 3, but exceptions are allowed if there are also strong [FeVII] and [FeX] emission lines present. FWHM(H$_{\beta}$) < 2000km/s (Goodrich 1989). In SDSS spectra they are often wrongly automatically classified as QSO because of FeII features seen as broad emission lines.\\

\textbf{S1h} (S1 hidden, defined in VCV-13). S1 showing S1 like spectra in polarized light (Antonucci $\&$ Miller 1985; Miller $\&$ Goodrich 1990; Tran et al. 1992). Seyfert 1 with a hidden BLR. \textbf{Hidden AGN} (X-ray revealed AGN not having emissions lines in optical range; Barger et al. 2000; Treister et al. 2005). These may be classified as X-ray AGN (XAGN; Paronyan $\&$ Mickaelian 2015). Among other AGN detected in optical wavelengths, very small fraction displays X rays (V$\acute{e}$ron-Cetty et al. 2004). \textbf{NELG (NLXG)} (Narrow Emission Line Galaxy, Narrow-Line X-ray Galaxy) are most likely obscured Seyfert galaxies, also some kind of hidden AGN.\\

To identify the known AGN and to establish the available types, we have also cross-correlated our sample with the Catalogue of QSO and AGN, 13th version (VCV-13, V$\acute{e}$ron-Cetty $\&$ V$\acute{e}$ron 2010), the Large Quasar Astrometric Catalogue (LQAC-2) (Souchay et al. 2012), SDSS-DR10 Quasar Catalog (P$\hat{a}$ris et al. 2014), Roma Multi-frequency Catalogue of Blazars (BZCAT) 5th version (Massaro et al. 2015), and Low-frequency radio catalog of flat-spectrum sources (FSS) (LORCAT, Massaro et al. 2014). All available databases (NED, HyperLEDA) have also been searched for object types. These results are given in \textbf{Table 2}.\\
\begin{table}[h]
\begin{center}
 \caption{Available data for HRC/BHRC extragalactic X-ray sources from AGN catalogues.}
  \begin{tabular}{| l | c | c | c | c | c |}
    \hline
\textbf{Catalogue} & \textbf{Year} & \textbf{Search radius $''$} & \textbf{AGN} & \textbf{Gal} & \textbf{Most important data} \\ \hline
VCV-13 & 2010 & 30 & 3352 & 0 & Activity types, M$_{abs}$ \\ \hline
LQAC-2 & 2012 & 5 & 3198 & 5 & Accurate positions \\ \hline
BZCAT v.5 & 2015 & 5 & 387 & 0 & Radio, X-ray, $\gamma$-ray \\ \hline
LFRC FSS & 2014 & 30 & 114 & 7 & Radio \\ \hline
SDSS DR12 spectro & 2015 & 2 & 2908 & 231 & 3800-9200 (3500-10500) AA \\ \hline
NED & 2015 & 30 & 4253 & 492 & Redshifts, morphology, etc. \\ \hline
HyperLEDA & 2015 & 30 & 1182 & 260 & Redshifts, activity types, etc. \\ \hline
\textbf{Total} & & & \textbf{4253} & \textbf{492} &  \\ \hline
  \end{tabular}
\end{center}
\end{table}

\section{Multiwavelength Data}
To have MW data for our sample, we cross-correlated the Joint HRC/BHRC sources with all-sky and large area catalogues in $\gamma$-rays (Fermi-GLAST 3FGL: Acero et al. 2015; INTEGRAL IBIS/ISGRI soft gamma-ray survey catalog: Bird et al. 2010), UV (GALEX AIS and MIS: Bianchi et al. 2011), optical range (APM: McMahon et al. 2000; MAPS: Cabanela et al. 2003; USNO-B1.0: Monet et al. 2003; GSC 2.3.2: Lasker et al. 2008; SDSS DR12 photometric and spectroscopic catalogues: Alam et al. 2015), NIR (2MASS PSC: Cutri et al. 2003 and 2MASS ESC: Skrutskie et al. 2006), MIR (WISE: Cutri et al. 2012; AKARI IRC Point Source Catalogue: Ishihara et al. 2010), FIR (IRAS PSC: IRAS 1988, IRAS FSC: Moshir et al. 1992 and IRAS SSSC: Helou et al. 1985, as well as we used Merged IRAS PSC/FSC: Abrahamyan et al. 2015; AKARI FIS Bright Source Catalogue: Yamamura et al. 2010), and radio (GB6: Gregory et al. 1996; NVSS: Condon et al. 1998; FIRST: Helfand et al. 2015; WENSS: de Bruyn et al. 1998; 7C: Hales et al. 2007).\\

\textbf{Table 3} gives MW data for X-ray sources from the Joint HRC/BHRC catalogue, as well as we have listed all photometric bands that may be used for further analysis and building MW SEDs. As a result, we have 41 photometric bands along very wide spectral range from 100 GeV in gamma-rays to 151 MHz in radio. However, because of limited sensitivity of many surveys, for most of objects only ROSAT, GALEX, APM/USNO/GSC, SDSS, 2MASS and WISE are available, resulting in 18 photometric bands.
\begin{table}[h]
\begin{center}
 \caption{Multiwavelength data for extragalactic X-ray sources from the Joint HRC/BHRC catalogue.}
  \begin{tabular}{| l | c | c | c | c | }
    \hline
\textbf{Catalogue} & \textbf{Search radius $''$} & \textbf{AGN} & \textbf{Gal} & \textbf{Photobands, spectral range} \\ \hline
Fermi-GLAST	& 600 & 155 & 5 & 0.1-0.3, 0.3-1, 1-3, 3-10, 10-100 GeV \\ \hline
INTEGRAL & 600 & 34 & 0 & 10, 30, 60 keV \\ \hline
ROSAT BSC/FSC & 40 & 4253 & 492 & 0.07-2.4 keV \\ \hline
GALEX & 3 & 2769 & 172 & FUV 1528A, NUV 2271A \\ \hline
APM & 4 & 4099 & 442 & b 4050, r 6452 \\ \hline
USNO-B1.0 & 4 & 4239 & 486 & B 4050/4680, R 6452, I 8060 \\ \hline
GSC 2.3.2 & 4 & 4213 & 466 & b$_{j}$ 4680, F 6452, N 8060 \\ \hline
SDSS DR12 photo & 2 & 3976 & 440 & u 3551, g 4686, r 6165, i 7481, z 8931 \\ \hline
SDSS DR12 spectro & 2 & 2908 & 231 & 3800-9200 (3500-10500) AA \\ \hline
2MASS & 2 & 3368 & 410 & J 1.24, H 1.66, Ks 2.16 $\mu$m \\ \hline
WISE & 2 & 4084 & 448 & w1 3.4, w2 4.6, w3 11.6, w4 22 $\mu$m \\ \hline
AKARI IRC & 10 & 71 & 1 & 9, 18 $\mu$m \\ \hline
IRAS PSC/FSC & 60 & 249 & 11 & 12, 25, 60, 100 $\mu$m \\ \hline
AKARI FIS & 5 & 81 & 1 & 65, 90, 140, 160 $\mu$m \\ \hline
GB6 & 240 & 824 & 97& 6 cm (4.83 GHz) \\ \hline
NVSS & 10 & 930 & 135 & 21 cm (1.4 GHz) \\ \hline
FIRST & 3 & 1021 & 132 & 21 cm (1.4 GHz) \\ \hline
WENSS & 60 & 444 & 75 & 49 cm (610 MHz), 92 cm (330 MHz) \\ \hline
7C & 75 & 278 & 43 & 198 cm (151 MHz) \\ \hline
\textbf{Total} & & \textbf{4253} & \textbf{492} & \textbf{100 GeV - 151 MHz} \\ \hline
  \end{tabular}
\end{center}
\end{table}
These MW data allow full understanding of these objects, as well as analysis of their colour-magnitude and colour-colour diagrams. \textbf{Fig. 1} gives such diagrams for SDSS, 2MASS and WISE data, showing reliable distinguishing of stars from galaxies. However, one can notice that most active stars (CVs and WDs) show evidence of having similar to AGN and other galaxies properties. Here all non-classified sources have been grouped into candidate stars, galaxies, and AGN.\\
\begin{figure}[h]
\caption{Multiwavelength colour-magnitude and colour-colour diagrams based on
SDSS, 2MASS and WISE to distinguish stars from galaxies.}
\centering
\includegraphics[width=0.9\textwidth]{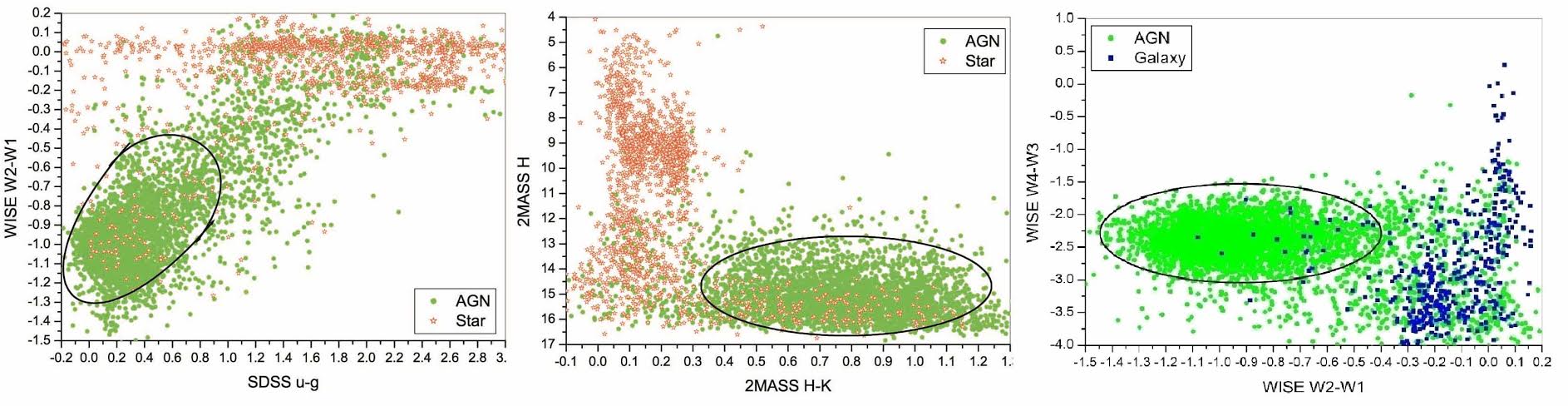}
\end{figure}

In \textbf{Fig. 2} we also use SDSS and 2MASS colour-magnitude diagrams to distinguish AGN and galaxies from stars and different types of extragalactic sources: QSOs, other AGN, and normal galaxies (possibly hidden AGN). We also use X-ray properties to find the stars against galaxies. X-ray and SDSS r flux ratio vs. X-ray flux diagram has been used.
\begin{figure}[h]
\caption{SDSS and 2MASS colour-magnitude diagrams to distinguish AGN and galaxies from stars and different types of extragalactic sources: QSOs, other AGN, and normal galaxies (possibly hidden AGN).}
\centering
\includegraphics[width=0.8\textwidth]{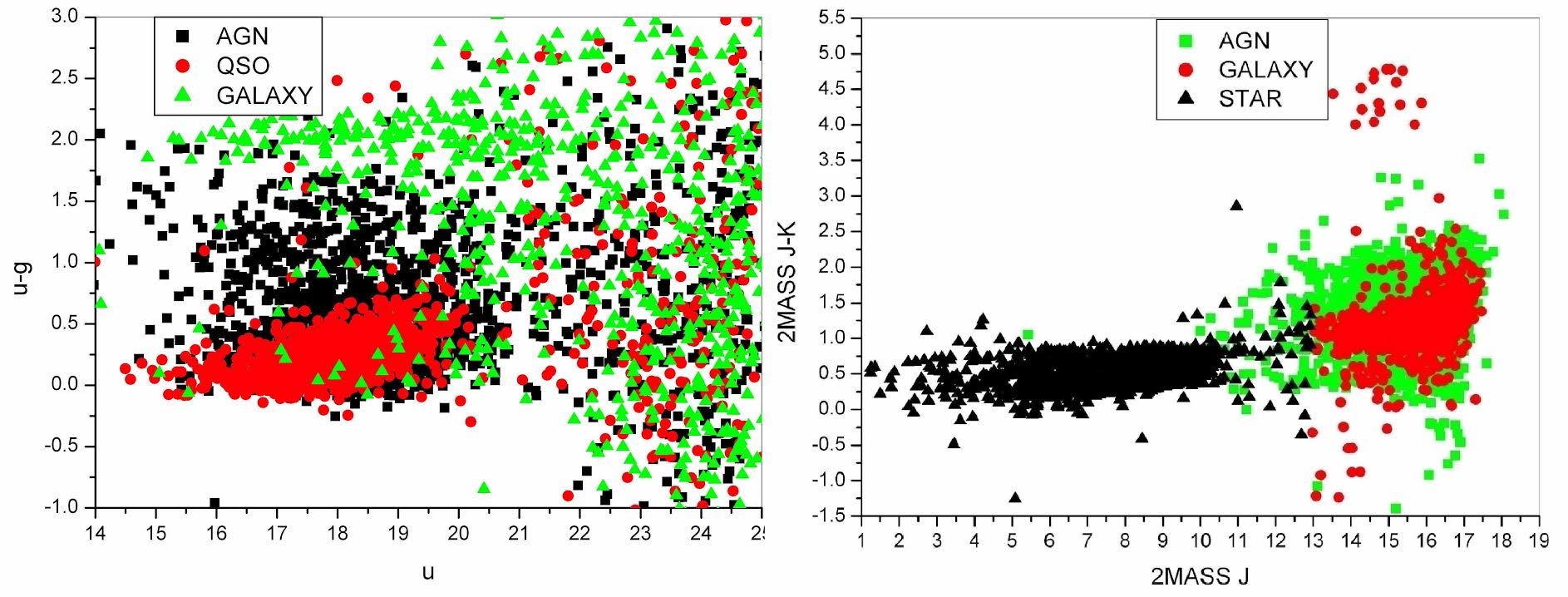}
\end{figure}

\section{Activity types}
We have retrieved SDSS DR12 spectra and carried out classification for activity types, as well as we have collected all available data on activity types from NED and HyperLEDA databases and AGN catalogues. We have used fine criteria for Seyferts, giving all subtypes (Sy1.0 - S2.0) and especially focusing on narrow line Seyferts (and QSOs), as these are typical X-ray sources. We particularly examined narrow broad lines and along with standard NLS1, we introduced NLS1.0, NLS1.2 and NLS1.5 subtypes (NLS1.8 and NLS1.9 are rather difficult to identify). Out of 884 AGN candidates, we have carried out classification for 173 ones. Given that only 61 were confirmed as genuine AGN, now we will have 3413 objects in AGN sample. Together with misclassified “Stars” sample objects, finally we have the distribution of types shown in \textbf{Table 4}.\\
\begin{table}[h]
\begin{center}
 \caption{Distribution of activity types for all HRC/BHRC extragalactic objects using SDSS DR12 spectra and other available data on activity.}
  \begin{tabular}{| l | c | c | }
    \hline
\textbf{Type} & \textbf{Number} & \textbf{Fraction ($\%$)}  \\ \hline
BL Lac + HPQ & 320+13 & 7.02 \\ \hline
QSO & 1309 & 27.59 \\ \hline
Sy1.0-1.5 & 1876 & 39.54 \\ \hline
Sy1.8-2.0 & 111 & 2.34 \\ \hline
Sy: & 37 & 0.78 \\ \hline
LINER & 21 & 0.44 \\ \hline
AGN & 82 & 1.73 \\ \hline
AGN: & 451 & 9.50 \\ \hline
Starburst / HII & 25 & 0.53 \\ \hline
Abs & 8 & 0.17 \\ \hline
Bright galaxy & 492 & 10.37 \\ \hline
\textbf{Total} & \textbf{4745} & \textbf{100.0} \\ \hline
  \end{tabular}
\end{center}
\end{table}

We give several examples of different objects based on our SDSS classification.

\subsection{Bright Galaxies}
Many bright galaxies show X-ray (as well as they radiate at all wavelength ranges) even without having any sign of activity. One of our tasks is to establish the limit of ROSAT detection of normal bright galaxies. Then all fainter objects may turn to be relevant AGN candidates even without having their spectra.

\subsection{QSOs}
QSOs are most typical and comprise the largest sample among extragalactic X-ray sources. In \textbf{Figure 3} we give two examples of HRC/BHRC QSOs revealed as QSO candidates based on HQS low-dispersion spectra and later on discovered by means of SDSS medium resolution spectra.
\begin{figure}[h]
\caption{SDSS DR12 spectra for 2 HRC/BHRC QSOs. On the left panel we observe typical to Sy1.5 feature (combination of broad and narrow emission lines), which allow us to classify this objects as QSO1.5.}
\centering
\includegraphics[width=0.8\textwidth]{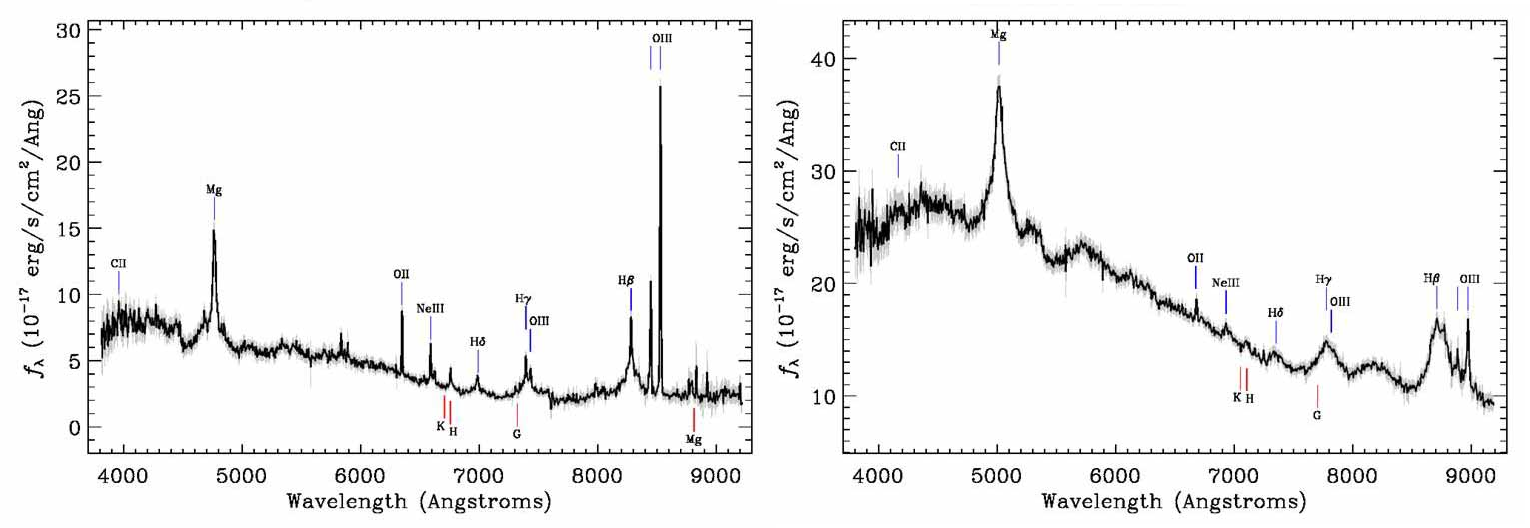}
\end{figure}

\subsection{BL Lacs}
Given that BL Lacs typically do not show any (emission or absorption) lines, it is rather difficult to distinguish them from stars (as very often stellar absorption lines are weak and not detectable). However, due to synchrotron radiation, their continua are linear compared to the characteristic Planck curve for the stars. Hence, only high quality spectra and very accurate reduction may reveal their spectral energy distribution and prove their nature even without measuring the redshift. On the other hand, presence of radio may also be a relevant criterion to distinguish BLL against blue stars.

\subsection{NLS1s}
Narrow Line Seyfert 1 galaxies are among the most interesting objects in our sample. We have found different types of such objects; NLS1.0, NLS1.2 and NLS1.5, as well as NLS1.8 and NLS1.9 may exist, however due to small intensity of their broad lines, only high signal-to-noise and high resolution spectra may reveal them. In \textbf{Figure 4}, two examples of NLS1s are given.
\begin{figure}[h]
\caption{SDSS DR12 spectra for 2 HRC/BHRC NLS1s. Many FeII lines are observed on both sides of H$_{\beta}$.}
\centering
\includegraphics[width=0.8\textwidth]{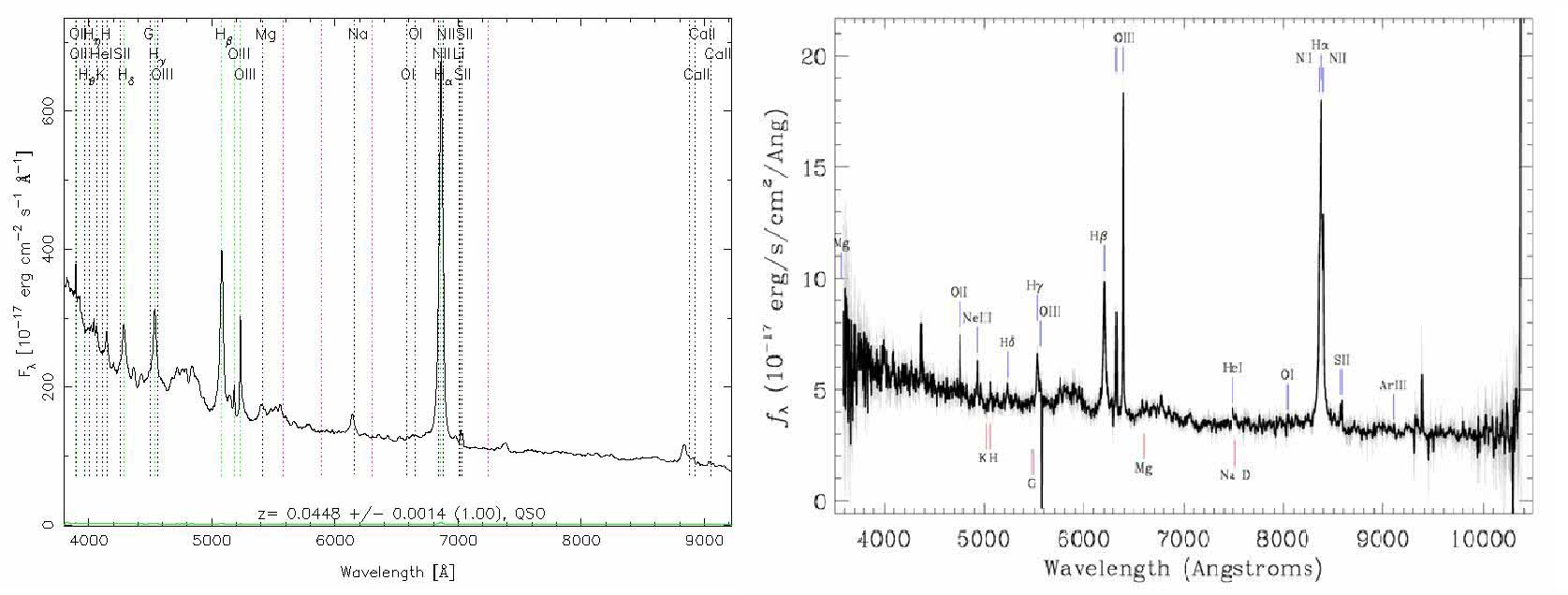}
\end{figure}

\subsection{Other AGN}
Other AGN include various types of Seyferts (from Sy1.0 to Sy2.0 in the order of AGN luminosity) and LINERs. We give in \textbf{Figure 5} SDSS spectra for 2 Seyfert galaxies; (left) Sy1.5 (broad and narrow components have similar intensity) and (right) Sy1.8 (broad components are weak and observed only for H$_{\alpha}$ and H$_{\beta}$ lines).
\begin{figure}[h]
\caption{SDSS DR12 spectra for 2 HRC/BHRC Seyfert galaxies: (left) Sy1.5 (broad and narrow components have similar intensity) and (right) Sy1.8 (broad components are weak and observed only for H$_{\alpha}$ and H$_{\beta}$ lines).}
\centering
\includegraphics[width=0.8\textwidth]{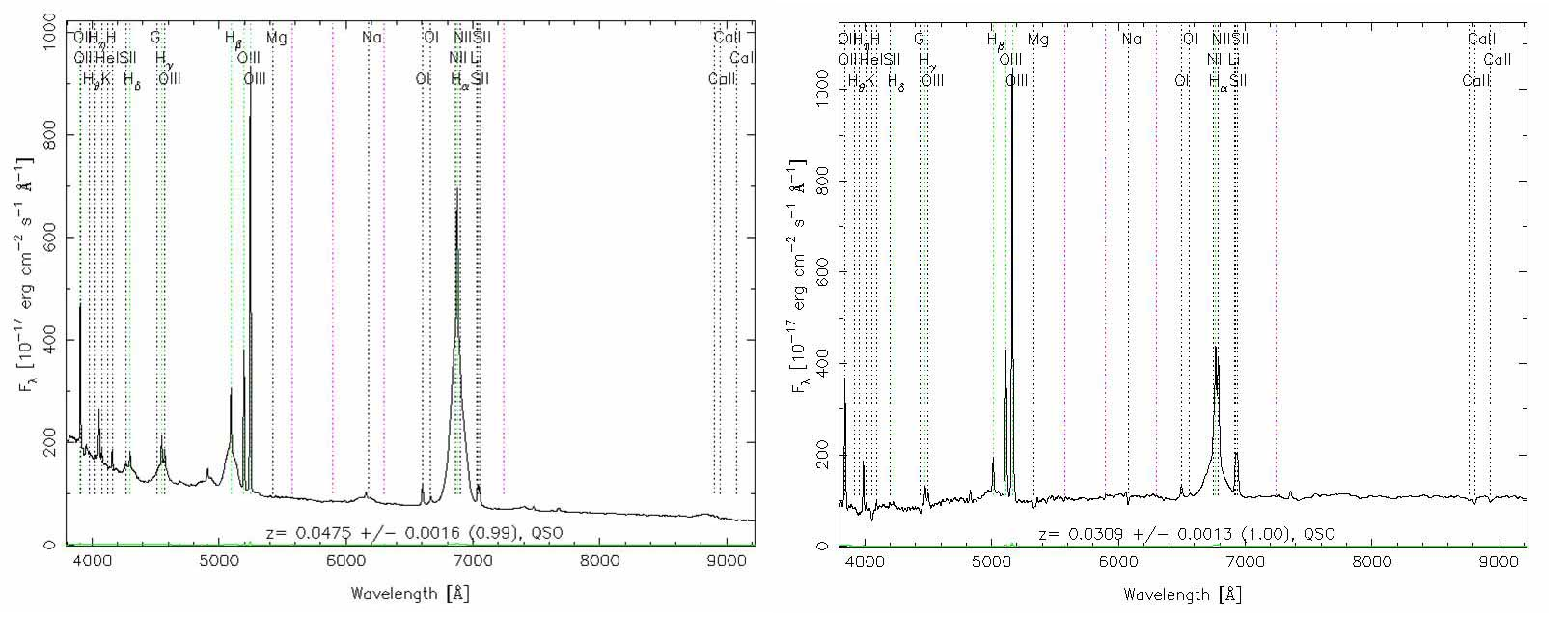}
\end{figure}

\subsection{Starburst or HII galaxies}
Starburst (or HII) galaxies have no AGN in the center, however due to undergoing intense star formation process (SFR is a few M$_{\odot}$ yr$^{-1}$, but may reach up to 103 M$_{\odot}$ yr$^{-1}$) and presence of heated gas they show strong Balmer and some forbidden emission lines (OIII, OII, OI, NII, SII and others. They may have enough strong stellar component, as well as strong continuum, which sometimes is very blue. In \textbf{Figure 6}, two examples of HII galaxies are given.
\begin{figure}[h]
\caption{SDSS DR12 spectra for 2 HII galaxies. The galaxy on right panel has very blue continuum.}
\centering
\includegraphics[width=0.8\textwidth]{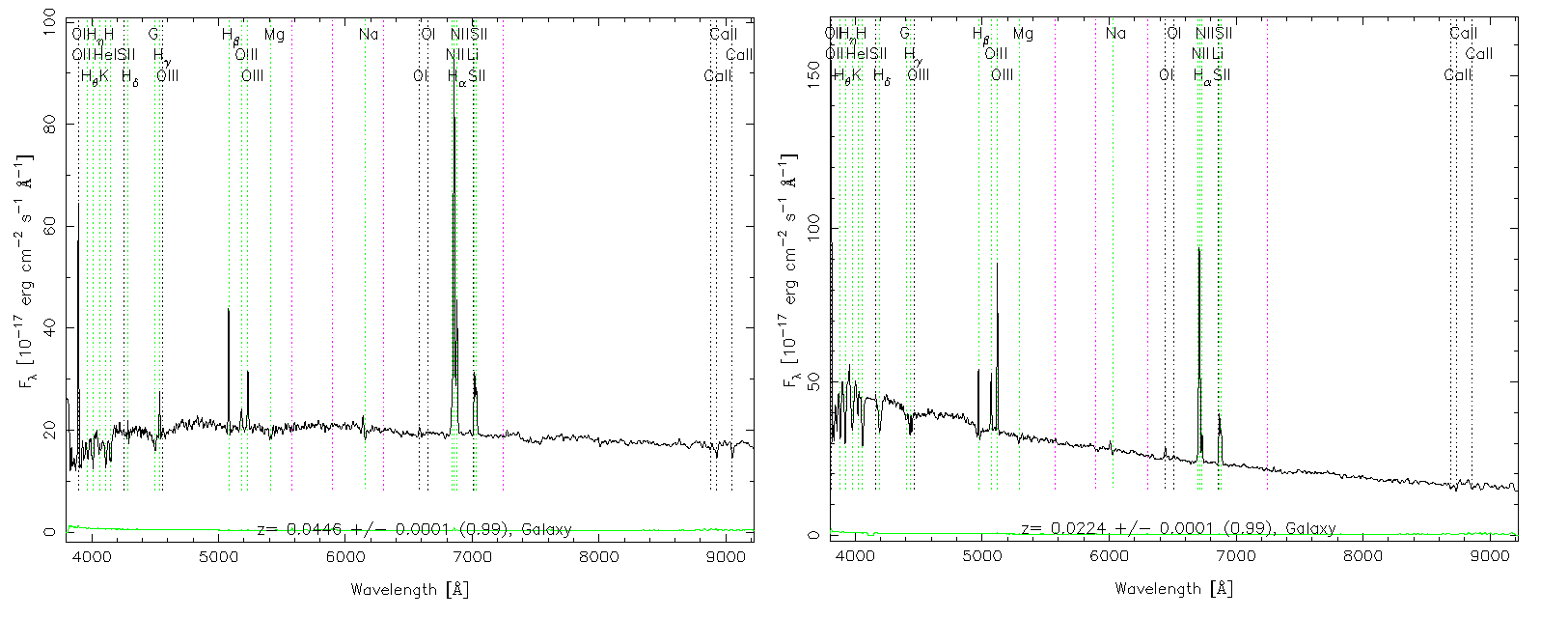}
\end{figure}

\subsection{Absorption line galaxies: hidden AGN}
Absorption line galaxies do not show any sign of activity when studies by means of optical spectra. However, the presence of strong X-ray and in many cases also radio suggests that they are hidden AGN, so that they are completely opaque in optical light, but they may be exactly the same AGN similar to classical Seyferts. We give in \textbf{Figure 7} two examples of such galaxies that are typically red ones.
\begin{figure}[h]
\caption{SDSS DR12 spectra for 2 absorption line galaxies typically showing Balmer, NaII and MgI absorption lines and having no sign of activity based on optical spectra.}
\centering
\includegraphics[width=0.8\textwidth]{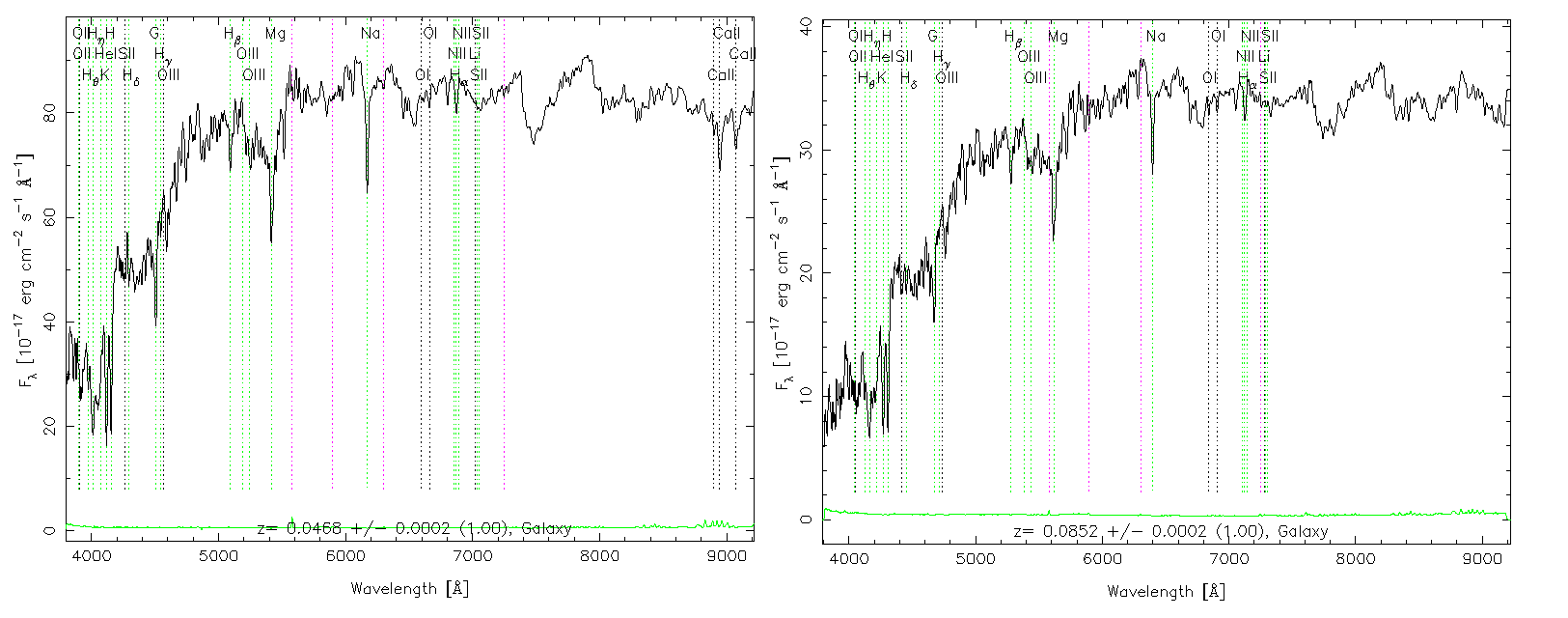}
\end{figure}

\subsection{Interacting Galaxies}
We have discovered red interacting galaxies that are strong X-ray sources. Most powerful extragalactic IR sources (ULIRGs and HLIRGs) are interacting/merging galaxies. However, the X-ray sample among such objects must be investigated in detail.

\subsection{Clusters of Galaxies}
Many remote clusters have been detected in ROSAT as single point sources. The question is to check if there are strong X-ray emitters among the cluster galaxies (e.g. central giant ones) or X-ray comes from the intergalactic gas in the cluster. New X-ray missions, such as XMM-Newton and Chandra may answer this question.

\subsection{Misclassified Objects: Stars}
As X-ray sources contain many interesting stars, some fraction of contamination is always present, especially because our former classification was based on low-dispersion spectra. Typical types of stars that appear to be X-ray sources are: white dwarfs (WD), cataclysmic variables (CV), late-type stars (K, M, C) and others. We give the results of SDSS spectral classification of these objects in another paper (Mickaelian et al. 2015).

\section{Summary and Future Studies}
A project on study of X-ray selected galaxies has been conducted in BAO. It includes selection, cataloguing, spectral classification, multiwavelength studies and statistical analysis of such objects.\\ \\
Follow-up programs and future studies based on X-ray selected AGN are:
\begin{itemize}
  \item Estimation of ROSAT BSC/FSC extragalactic content
  \item Optical variability of HRC/BHRC extragalactic sources (based on the analysis of DSS1/DSS2 photometric data from USNO-A2.0, APM, MAPS, USNO-B1.0 and GSC 2.3.2, as shown in Mickaelian et al. 2011),
  \item Radio 1400 MHz variability of HRC/BHRC extragalactic sources (based on NVSS-FIRST cross-matching and study of radio fluxes of associated sources),
  \item Detailed spectral study (planned, the results will be published soon),
  \item Creation of ROSAT-NVSS AGN sample (a preliminary list of 6,099 X-ray/radio sources without an optical identification; brighter objects are normal bright galaxies, while all faint ones are candidate AGN with some contamination of distant clusters),
  \item Study of X-ray selected blazars (to support their definition in terms of X-ray properties),
  \item Search for optical counterparts of all ROSAT sources (using available catalogues of bright stars and galaxies, AGN, WDs, CVs, late-type stars, etc. to distinguish different types of objects then using extension, radio, proper motion, optical colours, SDSS/2MASS/WISE diagrams for better understanding of the nature of these objects).
\end{itemize}
There still are a number of open questions related to X-ray selected AGN, such as:
\begin{enumerate}
  \item What is the fraction of X-ray AGN among all?
  \item Are there X-ray galaxies (like radio galaxies)?
  \item Is there a limit between X-ray luminosities between normal galaxies (background radiation) and AGN?
  \item Is there a correlation between optical and radio variability of AGN?
  \item What is the definition of blazars?
\end{enumerate}
The question of definition of blazars is still open. There are many parameters that may be regarded as such criteria: high luminosity, flat radio spectrum, strong X-ray and $\gamma$-ray, optical and/or radio variability, polarization, etc. Future studies will work on answering to these questions.


\begin{thebibliography}{}
\scriptsize {
\bibitem{}Abrahamyan, H. V.; Mickaelian, A. M.; Knyazyan, A. V. 2015, A$\&$C 10, 99
\bibitem{}Acero, F.; Ackermann, M.; Ajello, M.; et al. 2015, ApJS 218, 23
\bibitem{}Alam, S.; Albareti, F. D.; Allende Prieto, C.; et al. 2015, ApJS 219, 12
\bibitem{}Antonucci, R. R. J.; Miller, J. S. 1985, ApJ 297, 621
\bibitem{}Bianchi L.; Herald J.; Efremova B.; et al. 2011 ApSS 335, 161
\bibitem{}Bird A. J., Bazzano A., Bassani L., et al. 2010 ApJSS 186, 1
\bibitem{}Cabanela, J. E., Humphreys, R. M., Aldering, G., et al.: 2003 PASP 115, 837
\bibitem{}Condon, J. J., Cotton, W. D., Greisen, E. W.; et al.: 1998 AJ 115, 1693
\bibitem{}Cutri R. M.; Wright E. L.; Conrow T.; et al. 2012, WISE All-Sky DR, VizieR Catalog II/311
\bibitem{}Cutri, R. M.; Skrutskie, M. F.; van Dyk, S.; et al. 2003, IPAC/California Institute of Technology
\bibitem{}de Bruyn G., Miley G., Rengelink R., et al. 1998, WENSS Collab. NFRA/ASTRON and Leiden Obs.
\bibitem{}Gregory, P. C., Scott, W. K., Douglas, K., Condon, J. J. 1996, ApJS 103, 427
\bibitem{}Hagen, H.-J.; Groote, D.; Engels, D.; Reimers, D. 1995, A$\&$AS 111, 195
\bibitem{}Hales, S. E. G.; Riley, J. M.; Waldram, E. M.; et al. 2007, MNRAS 382, 1639
\bibitem{}Helfand, D. J.; White, R. L.; Becker, R. H. 2015, ApJ 801, 26
\bibitem{}Helou, G.; Walker, D. W. 1985, IRAS small scale structure catalog, JPL, Pasadena, NASA IRAS, 1988, Joint IRAS Scienve Working Group. IRAS PSC, Version 2.0, NASA RP-1190
\bibitem{}Ishihara, D.; Onaka, T.; Kataza, H.; et al. 2010, A$\&$A 514, 1
\bibitem{}Lasker, B. M.; Lattanzi, M. G.; McLean, B. J.; et al. 2008, AJ 136, 735
\bibitem{}Massaro, F.; Giroletti, M.; DAbrusco, R.; et al. 2014, ApJS 213, 3
\bibitem{}Massaro, E.; Maselli, A.; Leto, C.; et al. 2015, Ap$\&$SS 357, 75
\bibitem{}McMahon, R. G.; Irwin, M. J.; Maddox, S. J.; 2000, IoA, Cambridge, UK
\bibitem{}Mickaelian A. M., Hovhannisyan L. R., Engels D., Hagen H.-J., Voges W. 2006, A$\&$A 449, 425
\bibitem{}Mickaelian A.M., Mikayelyan G.A., Sinamyan P.K. 2011, MNRAS 415, 1061
\bibitem{}Mickaelian, A. M.; et al. 2015, AApTr, in press
\bibitem{}Miller, J. S.; Goodrich, R. W. 1990, ApJ, 355, 456
\bibitem{}Monet D. G., Levine S. E., Casian B., et al. 2003, AJ 125, 984
\bibitem{}Moshir, M., Kopan, G., Conrow, T., et al. 1990, IRAS FSC, Version 2.0, NASA
\bibitem{}Osterbrock, D. E.; Pogge, R. W. 1985, ApJ 297, 166
\bibitem{}P$\hat{a}$ris, I.; Petitjean, P.; Aubourg, E.; et al. 2014, A$\&$A 563, 54
\bibitem{}Paronyan, G. M.; Mickaelian, A. M. 2015, Ap$\&$SS, in press
\bibitem{}Schneider D.P., Richards G.T., Hall P.B., et al. 2010, AJ 139, 2360
\bibitem{}Schwope, A., Hasinger, G., Lehmann, I., et al. 2000, AN, 321, 1
\bibitem{}Skrutskie, M. F.; Cutri, R. M.; Stiening, R.; et al. 2006, AJ 131, 1163
\bibitem{}Souchay J., Andrei A.H., Barache C., Bouquillon S., Suchet D., Taris F., Peralta R. 2012, A$\&$A 537, A99
\bibitem{}Tran, H. D.; Osterbrock, D. E.; Martel, A. 1992, AJ, 104, 2072
\bibitem{}V$\acute{e}$ron-Cetty M.P., V$\acute{e}$ron P. 2010, A$\&$A 518, A10
\bibitem{}V$\acute{e}$ron-Cetty, M.-P.; Balayan, S.K.; Mickaelian, A.M.; et al. 2004, A$\&$A 414, 487
}
\end{thebibliography}
\end{document}